\documentclass[a4paper,10pt]{article}   

  \usepackage{makeidx}
  \usepackage{graphicx,graphics,color}
  \usepackage{latexsym,doc,fontenc,exscale}
\usepackage{amsmath,amsfonts,amsthm,eucal,amssymb}

\begin{document}

 \title{\hspace{-0.5cm}\bf Supersymmetry Across Nanoscale Heterojunction}

\author{\textbf{B.~Bagchi}$^{1}$\footnote{bbagchi123@rediffmail.com}, \textbf{A.~Ganguly}$^{2}$\footnote{gangulyasish@rediffmail.com}
 and \textbf{A.~Sinha}$^{1}$\footnote{anjana23@rediffmail.com}\\
$^{1}$\small{Department of Applied Mathematics, University of Calcutta},\\ \small 92 Acharya Prafulla Chandra Road, Kolkata-700009, India\\
 $^{2}$\small City College, University of Calcutta, \\\small 13 Surya Sen Street,
 Kolkata-700012, India}
\maketitle
 \begin{abstract}
We argue that supersymmetric transformation could be applied
across the heterojunction formed by joining of two mixed
semiconductors. A general framework is described by specifying the
structure of ladder operators at the junction for making
quantitative estimation of physical quantities. For a particular
heterojunction device, we show that an exponential grading inside
a nanoscale doped layer is amenable to exact analytical treatment
for a class of potentials distorted by the junctions through the
solutions of transformed Morse-Type potentials.

 \vspace{2mm}
 \noindent
 \hspace{-.8cm}PACS number(s): 03.65.-w,03.65.Ca,0.65.Ge,73.40.Kp,73.40.GK,73.40.Ty,73.43.Jn,02.30.Hq,02.30.Gp
 \end{abstract}


To have a proper grasp of designing any semiconductor device that
relies on a heterojunction, a complete understanding of the
transport phenomena at interface is important \cite{koh}. Positive
and negative carriers moving across a junction \cite{as9,ag9} face
an abrupt jump in potential due to the periodicity defects in
crystal lattice at interface and varying electrical and optical
properties of different materials bounding the heterojunction.
Different values of the position-dependent conduction-band energy
on the two sides separated by a junction results in a
$\delta$-like singularity at the interface \cite{zhu} producing
either a very deep, narrow well or a very high but a thin barrier
that depends on the position of the junction. In these situations
it is worth considering the transformed potential which in a
supersymmetric context shares almost the same bound spectra but
are free from such a singularity. A general version of
supersymmetric transformation (see the monograph \cite{bagbook}
for an up-to-date survey) allows to construct a hierarchy of
potential chain which, in physical purpose, is useful to remove or
embed arbitrary number of bound states. Since the position of a
bound state is related to the pole-structure of the S matrix, its
artificial designing can be utilized to improve the mobility of
the carriers and thus having a desired output at the acceptor
diode. A study of the spectral properties of the induced
intertwined potentials across the heterojunction is thus of
immense physical importance.

In actual microscopic description of the motion of electrons (or
holes with $e\rightarrow -e$) across a semiconductor junction, it
is known \cite{kan} that the supersymmetric structure of the
underlying Dirac Hamiltonian describes unpaired `spin-up' and
`spin-down' states in the conduction and valence band near
$\Gamma$ or L-point in the Brillouin zone that are localized at
the junction. Note that the existing literature [7-11] does not
apply to a practical semiconductor device where two graded mixed
semiconductors are artificially glued (say Ga$_{1-x}$Al$_x$As/GaAs
heterojunction or Si-SiO$_2$ layer) since the role of ladder
operators across the junction is unclear. Consequently the usual
techniques of constructing the transformed potential fail in the
effective-mass approximation which replaces actual wave function
by envelope wave function. In this Report we show that the
information about spectral properties of a given effective mass
Hamiltonian having a $\delta$-singularity at the heterojunction
can be extracted from the supersymmetrically transformed
Hamiltonian which is free from such a singularity. We apply our
method for a particular Ga$_{1-x}$Al$_x$As/GaAs semiconductor
device to compute analytically local electron density based on
inter subband energy levels and envelope wave functions obtained
from a physically justifiable choice of grading function.

We start by considering two heterojunctions formed by an
artificial joining of three  different material layers (doped or
pure). In the effective mass approximation the normalized envelope
wave function of a single electron in the n-th subband moving
along the normal to the surface of the material layer (xy-plane)
obeys the Schr\"odinger equation~\cite{ben} for the given
Hamiltonian $H_0(z)$ in SI unit
\begin{equation}\label{se}
    H_0(z)\psi_n(z)\equiv \left ( -\frac{d}{dz}\left [\frac{\hbar^2}{2m^*(z)}\frac{d}{dz}\right ]+V_0(z)\right )\psi_n(z)=E_n\psi_n(z)
\end{equation}
%
The effective mass $m^*$ has a position-dependent variation
$m_2(z)$ between two junctions at $z=a_1,a_2$ connecting
continuously two known constant values $m_1$ and $m_2$ outside the
junctions
\begin{equation}\label{gm}
    m^*(z)=\sum_{j=1}^{3}G_j m_j\, ; \qquad m_2(a_i)=m_i\, ,\quad i=1,2\, .
\end{equation}
In above equations $G_j$s are dimensionless Heaviside theta
functions defined by the relations
 \begin{equation}\label{theta}
    G_1=\Theta (\frac{a_1-z}{a_0})\, , \quad G_3=\Theta (\frac{z-a_2}{a_0})\, , \quad  G_2=1-(G_1+G_3),
\end{equation}
where $a_0=4\pi\epsilon_0\hbar^2/m_0e^2=0.0529\mbox{ nm}$ is Bohr
radius, $\epsilon_0$ denoting permittivity of free space and
$m_0,e$ are free electron mass and charge.
 A general class of the potential $V_0(z)$ containing the
$\delta$-singularity at the junctions is
\begin{equation}\label{gp}
 V_0(z)=\!\sum_{j=1}^{3}G_j V^{(j)}_0-
  \frac{\hbar^2m_2^{'}}{a_0m_2^{2}}\sum_{\ell=1}^{2}\epsilon_{\ell} G_{\ell}G_{\ell+1}\delta \left (\epsilon_{\ell} \frac{z-a_{\ell}}{a_0}\right )\, ,
  \qquad \left [ m_2'\equiv \frac{dm_2}{dz}\, , \epsilon_{\ell}=(-1)^{\ell} \right ]\, .
\end{equation}
The actual form of the component potentials $V^{(j)}_0(z)$ will
depend on chemical composition of materials in different regions
and the applied electric and magnetic field.

We assume that inside each region (except for the junctions) the
 component Hamiltonians $H_0^{(j)}=-d/dz[(\hbar^2/2m_j)d/dz]+V^{(j)}_0$ possess
 formal solutions $\psi_{E}^{(j)}(z)$ for an arbitrary energy $E$. We
 shall prove that there exists transformed Hamiltonian
 $\widetilde{H}_0(z)$ that shares the same spectra as  $H_0$ but
 is free from $\delta$-singularity. In other words using supersymmetric factorization we are going to
construct a new  potential of the form
\begin{equation}\label{tp}
   \widetilde{V_0}(z)=\sum_{j=1}^{3}G_j\widetilde{V_0}^{(j)}(z),
\end{equation}
 satisfying the above properties where $\widetilde{V_0}^{(j)}(z)$ are non-singular. Note that the component Hamiltonians can always be
 projected in a factorized form
\begin{equation}\label{ff}
    H_0^{(j)}=A_jA^{\dagger}_j-\mathcal{E},\quad A_j=\frac{\hbar}{\sqrt{2m_j}}\frac{d}{dz}+W(z),
\end{equation}
where the superpotential $W$ is a continuous smooth function
determined by the shape of the component potentials. Our aim is to
factorize the given Hamiltonian $H_0$ in terms of the ladder
operators which will act globally through the junction,
i.~\hspace{-3pt}e.\ $H_0=A A^{\dagger}-\mathcal{E}$ (factorization
energy $\mathcal{E}$ is chosen to make spectra positive definite).

Since all junctions are quantum mechanically of equal preference
it is reasonable to choose the form of the ladder operators at
each junction as the average of the two adjacent
 component operators. This leads to the following representation of $A$:
\begin{equation}\label{gr}
  A=\sum G_j A_j=\frac{\hbar}{\sqrt{2m^*}}\frac{d}{dz}+W(z)
\end{equation}
 It is straightforward to verify that the presence of theta functions in the ladder
 operators simulates $\delta$-singularity in the given potential
 $V_0(z)$. It remains to show that the
transformed Hamiltonian given by the factorized form
\begin{equation}\label{th}
    \widetilde{H}_0(z)=A^{\dagger}A-\mathcal{E}
\end{equation}
will not contain such a singularity. Clearly the wave function
$\psi$ of the given Hamiltonian (\ref{se}) must be continuous in
the entire region for otherwise Schr\"odinger equation (\ref{se})
has to hold a stronger singularity than $\delta$. Noting that the
intertwining relation
\begin{equation}
    H_0A=A\widetilde{H_0}
\end{equation}
 relates $\psi$ and $\tilde{\psi}$ as $\psi \propto A\tilde{\psi}$,
we conclude that the action of $A$ on $\tilde{\psi}$ must be
continuous across junctions. Suppose that the the wave functions
 $\tilde{\psi}$ of $\widetilde{H}_0$ may be
constructed as a superposition of component solutions
$\tilde{\psi}_E^{(j)}$ where the average of two adjacent plain
waves contributes equally at the junction. Using this construction
and the representation (\ref{gr}) of the operator $A$ we then
arrive at the following matching conditions on $\tilde{\psi}$
\begin{equation}\label{match}
    \tilde{\psi}'(a_j+0)=\tilde{\psi}'(a_j-0)=\tilde{\psi}'(a_j).
\end{equation}
where we have used the usual continuity condition for
$\tilde{\psi}$. The smoothness of $\tilde{\psi}$ across junction
indicates that transformed potential (\ref{tp}) is free from any
singularity. This can be directly verified from the explicit
expression of the transformed potential
\begin{equation}\label{tpf}
    \widetilde{V_0}(z)=W^2-(\frac{\hbar W}{\sqrt{2m^*}})'-\mathcal{E}\, .
\end{equation}
If one applies similar reasoning for $\tilde{\psi}\propto
A^{\dagger}\psi$, a slope-discontinuity of $\psi$ will appear at
the junctions which is the signature of $\delta$-singularity. This
striking difference between two transformed Hamiltonians arises
due to the non-commutativity of momentum and mass operators which
we can exploit via supersymmetric transformation.

Equations (\ref{gr}),(\ref{match}) and (\ref{tpf}) constitute the
central result of this Report establishing a way to apply
supersymmetric transformation across heterojunction. The set of
conditions in (\ref{match}) yield a transcendental energy equation
whose roots correspond to the allowed electronic energy levels for
the given Hamiltonian except for the factorization energy
$\mathcal{E}$. Finally  we note that for practical application of
above result the particular composition of material layers will
guide the choice of the superpotential simulating the potential
structure inside and outside the junction.

Let us consider a device made up of Ga$_{1-x}$Al$_{x}$As/GaAs with
the interface located at $z=0$. The n-type Ga$_{1-x}$Al$_{x}$
layer is grown in two Regions I($-z_h<z<-a_t$) and
II($-a_t<z<a_t$) and adjacent GaAs layer is grown in Region
III($a_t<z<z_t$) so that two heterojunctions are symmetrically
placed about the origin. The potential $V_0(z)$ is generally
formed from several contributions
\begin{equation}\label{fullpot1}
V_0(z)=-e\phi(z)+V_h(z)+V_{\textrm{Im}}(z)+V_{\textrm{xc}}(z)
\end{equation}
 The electrostatic potential $\phi(z)$ satisfies Poisson's equation (SI unit)
\begin{equation}\label{poisson}
 \frac{d}{dz}\left [ \epsilon_0\kappa(z)\frac{d\phi}{dz}\right ]=e \, n(z),
\end{equation}
where $n(z)=\sum_n N_n\psi_n^2(z)-N_D(z)+N_{AC}(z)$ is the
electron concentration in which $N_D (z),N_{AC}(z)$ denote the
position-dependent donor and acceptor concentrations and $N_n(z)$
represents the number of electrons per unit of area (in units of
cm$^{-2}$)in subband n: $N_n(z)=(m^*(z)k_BT/\pi \hbar^2)\ln
[1+\exp\{(E_F-E_n)/k_BT\}$, $E_F$ denoting Fermi energy. Since
electrostatic potential has to be extrapolated between (\ref{se})
and (\ref{poisson}) in an iterated numeric procedure, we shall not
consider it here. $V_{\textrm{Im}}(z)$ and $V_{\textrm{xc}}(z)$
are respectively image potential [arising due to the dielectric
constant step $\kappa (z)$] and exchange-correlation potential due
to electron-electron interaction in the channel.  It is well-known
that image potential cannot be incorporated in equation (\ref{se})
as it involves singularity at the origin. This effect is usually
realized into $V_{\textrm{xc}}(z)$ by adding an image term
dependent on mutual coordinates of two interacting electrons in
the population. However for $\kappa_b\approx\kappa_c$ [the
subscripts $b$ and $c$ indicate the barrier (Ga$_{1-x}$Al$_x$As)
and the channel (GaAs) sides of heterojunctions], image term could
be ignored and exchange-correlation potential may be parameterized
\cite{stern} as (in Hartree atomic unit of energy:
$\textrm{E}_{\textrm{h}}=\hbar^2/m_0a_0^2\simeq 27.2$ eV)
\begin{equation}\label{exchange}
 V_{\textrm{xc}}(z)=-\frac{1}{2}K_{cr}^2(x)\frac{2\ell m^*(z)/m_0}{\kappa^2(z)}\: \textrm{E}_{\textrm{h}},
\end{equation}
where the dimensionless constant $K_{cr}(x)$ depends on the doping
fraction $x=x(z)$ as
\begin{equation}\label{df}
    K_{cr}(x)=\sqrt{\frac{1+0.7734x\ln
    (1+x^{-1})}{21\ell\pi x(4/9\pi)^{1/3}}}
\end{equation}

 For the analytical purpose, a suitable continuous functional form for
$\kappa(z)$ is useful which connects two known values $\kappa_b$
and $\kappa_c$ of dielectric constant. A simple choice
$\kappa(z)=\sqrt{2\ell m^*(z)/ m_0}$ reduces $V_{xc}(z)$ to an
$x$-dependent constant simulating doping-influenced shift in the
energy scale where $\ell$ is a parameter adjusting with
$\kappa_{b}$ and $\kappa_{c}$. Usually in numerical process a
linear interpolation is taken between the values in the barrier
and channel side. We propose an exponential grading in the
intermediate Region II to connect values in I and III. This is
reasonable since in a sufficiently narrow thickness of layer II
($2a_t$ in unit of nm) it basically implies linear interpolation
and add corrections with the increase of thickness. Hence
following connection between known effective mass values
$m_b=m_2(-a_t)=bm_0$ and $m_c=m_2(a_t)=cm_0$ on Region I and III
will be adopted
\begin{equation}\label{mass}
    m^*(z)=G_1m_b+G_2 m_2+G_3 m_c \, , \qquad m_2=m_0\beta^2e^{2\beta z/a_0}
\end{equation}
where the dimensionless parameter $\beta$ is related with
mass-steps $b,c$ on both sides and the grading constants are given
by (\ref{theta}) with the replacement of $a_{1,2}$ by $\mp a_t$.
The choice of the superpotential will be guided by the physical
fact that Aluminium doping creates a barrier \cite{fey} for
electrons to diffuse to Region I and so they are mobilized to
acceptor site containing a GaAs quantum well. Thus the potential
$V_0$ should have a barrier on left, a well at intermediate region
and approaching a constant value on the right to make the
electrons pass through. This situation can be realized by the
following superpotential
\begin{equation}\label{sup-pot}
    W(z)=\frac{\hbar}{a_0\sqrt{2m_0}}(\mathcal{A} -\mathcal{B}e^{-\beta z/a_0}).
\end{equation}
\begin{figure}[ht]
 \begin{center}
 \includegraphics[height=11cm,width=15cm]{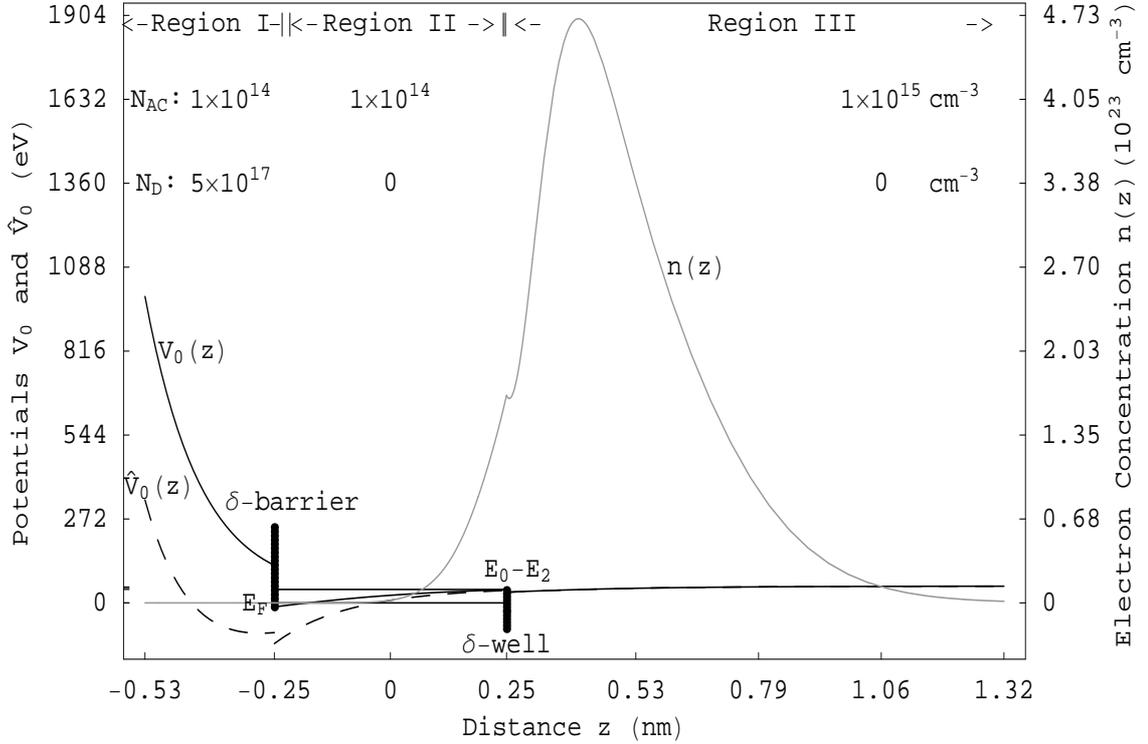}
 \end{center}
 \caption{\small{Original and transformed potentials  with three levels $E_0=43.58,E_1=46.75$ and $E_2=51.67$ eV ($E_F=35.9$ meV at 300 K) for the choice
 $\mathcal{A}=2,\mathcal{B}=1.1,\beta=0.21$ and $a_t=4.73$ which corresponds to 0.5nm intermediate layer thickness. Electron concentration is calculated
 using the data indicated in figure based on three wave functions obtained. Electrostatic term is not included.}}\label{fig1}
 \end{figure}
%

Both parameters $\mathcal{A}$ and $\mathcal{B}$ are dimensionless.
The latter parameter is free  controlling the depth of well and
the height of the barrier and the former is related with
exchange-correlation term (\ref{exchange}) through the definition
$\mathcal{A}=\sqrt{K^2-K_{cr}^2},K\geq K_{cr}$. At critical value
$K$$=K_{cr}$, the doping-influence will be prominent, since it has
very small values for $x=0.3$ and $0.4$. Our strategy is to solve
the transformed potential (\ref{tp}) according to the matching
condition (\ref{match}).

The add-on advantage of the model-choice (\ref{mass}) and
(\ref{sup-pot}) is that the component potentials
$\widetilde{V_0}^{(j)}$ in three regions are of Morse type and so
exact analytical solutions are available. The final expressions of
unnormalized solutions for the transformed Hamiltonian  are
\begin{equation}\label{ts}
  \left . \begin{array}{lcl}
 \tilde{\psi}^{(1)}_{E}(z)&=&\exp [-\surd b\{\chi (z)+\surd m_0\lambda z \}]U(g_b,h_b,2\surd b\chi)\vspace{3pt}\\
  \tilde{\psi}^{(2)}_{E}(z)&=&\exp [\frac{\mathcal{B} \beta z}{a_0}-y(z)][M(g_2,h_2,2y(z)+\varrho \, U(g_2,h_2,2y(z))]\vspace{3pt}\\
    \tilde{\psi}^{(3)}_{E}(z)&=&\exp [-\surd c\{\chi (z)+\surd m_0\lambda z\} ]M(g_c,h_c,2\surd c\chi)
           \end{array}  \right \}
\end{equation}
 where $\chi(z)\!=\mathcal{B} e^{-\beta z/a_0}/\beta$, $y(z)=a_0\lambda \surd m_0 e^{\beta z/a_0}$,
 $\lambda (E)=\surd [\mathcal{A}^2/m_0a_0^2-2(E+\mathcal{E})/\hbar^2]$
and $U(g,h,\xi)$, $M(g,h,\xi)$ are two linearly independent
solutions \cite{abr} of confluent hypergeometric equations
\begin{equation}\label{ch}
    \xi\frac{d^2w}{d\xi^2}+(h-\xi)\frac{dw}{d\xi}-gw=0
\end{equation}
with the arguments $g_j=\surd j(a_0\surd m_0\lambda
-\mathcal{A})/\beta,h_j=1+2a_0\lambda\sqrt{m_j} /\beta; j=b,c$ and
$g_2=(2\mathcal{B}-1)(\lambda-\mathcal{A})/2\lambda,h_2=2\mathcal{B}-1$.
The coefficient $\varrho$ in the second equation of (\ref{ts}) is
evaluated at solutions of energy equation. The normalized envelope
wave function of original Hamiltonian~(\ref{se}) for $n$-th
subband $E_n(>0)$ is
\begin{equation}\label{os}
   \psi_n(z)= \mathcal{N}_n(A\tilde{\psi}_{n+1}(z)),\quad
   \tilde{\psi}_n(z)=\sum_{j=1}^3G_j\tilde \psi^{(j)}_{E_n}(z).
\end{equation}
\begin{figure}[ht]
 \begin{center}
 \includegraphics[height=11cm,width=15cm]{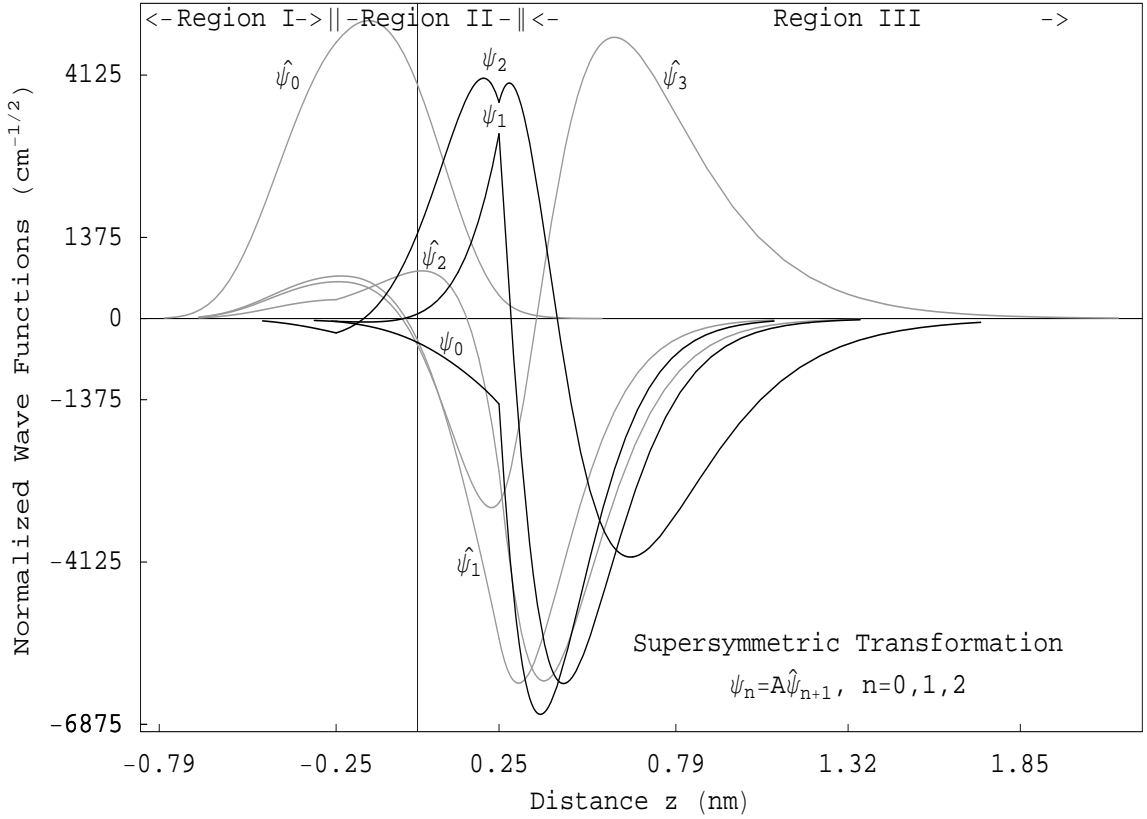}
 \end{center}
 \caption{\small{Envelope wave functions for the original and transformed potentials. Wave function for given potential
 is obtained by applying ladder operator $A$ which suppresses one node. Electrostatic term is not included.}}\label{fig2}
 \end{figure}

The energy equation is solved, using Mathematica, for
representative values: $\mathcal{A}=2$, $\mathcal{B}=1.1$ and
$\beta=0.21$ with the intermediate layer thickness 0.5nm.
Fig~\ref{fig1} shows the given potential (solid line) with
$\delta$-singularity and transformed potential (dashed line) with
$\delta$-singularity removed having four levels with zero energy
ground state. Thus we choose $\mathcal{E}=0$ and the three levels
for given potential are shown. The levels are very close
($E_{10}=3.17,E_{21}=4.92$ eV) and lie at the bottom of the well
above the Fermi energy $E_F=35.9$ meV at 300K. Electron
concentration (grey curve) is shown at 300K by choosing smooth
continuous functional form around small intervals containing two
junctions. Fig \ref{fig2} shows the effect of supersymmetric
transformations on wave functions $\tilde{\psi}_{n+1},n=0,1,2$
(Grey curves) which suppress one node to recover original wave
functions $\psi_n$, as is expected. Higher levels for transformed
potential violate oscillation theorem as $\delta$-singularities
resist creation of node near junctions. Interestingly original
wave functions perfectly follow oscillation theorem leaving a
sharp non-smoothness at the junction near channel side. Hence
transformed Hamiltonian, which is free from $\delta$-singularity,
digests its effects and makes the original solutions physically
acceptable.

To conclude, we have provided a method to bypass the essential
$\delta$-singularity in the effective-mass potential at the
heterojunction using supersymmetric transformation. Note that the
method is initiated from the observation that although original
potential is invariably affected by $\delta$-singularity at the
heterojunction, its supersymmetric partner escapes such a
singularity. We emphasize that this observation was unnoticed in
the literature. Further, in spite of extensive works [7-11]
regarding application of supersymmetry in position-dependent mass
problems, no prescription existed to tackle the presence of
heterojunction \cite{as9,ag9} in semiconductor. This gap is filled
in the present communication. Finally, for a particular
Ga$_{1-x}$Al$_{x}$As/GaAs device we have provided an exact
analytical treatment based on a model grading function connecting
values of band parameters across the junction. The choice of
exponential grading function is subtle for exact analytical
solution as it leads to a mapping in a constant-mass scenario to a
known Morse potential in Region~I and III which are joined by a
distorted Morse-Type potential in the intermediate Region~II.

We thank Dr.~S.~Ghosh for his interest in this work. AS is
supported by DST, Govt.\ of India.

\end{document}